# Evaluation of websites of state public health agencies during the COVID-19 pandemic demonstrating the degree of effort to design for accessibility

Arunkumar Pennathur[1*]

Amirmasoud Momenipour[2]

Priyadarshini Pennathur[1]

Brandon Murphy[1]

[1]Industrial and Systems Engineering Department, University of Iowa, Iowa City, IA 52246.

[2]Department of Engineering Management, Rose-Hulman Institute of Technology, Terre Haute, IN 47803.

*Author for correspondences. Email: arunkumar-pennathur@uiowa.edu

**Abstract**

Since the beginning of the pandemic, every state public health agency in the United States has created and maintained a website dedicated to COVID-19. Our goal was to evaluate these websites for conformity to accessibility guidelines. Specifically, we assessed, on a scale of increasing levels of accessibility compliance requirements, the results of the efforts made by website developers to incorporate and meet accessibility compliance criteria. We focused on homepages and vaccine pages in 49 states. For this study, we used the automated AChecker tool to assess conformance to the WCAG 2.0 guidelines at A, AA and AAA levels of conformance, and conformance with the Section 508c standard. We also manually rated, on a scale 0 (none) to 3 (highest), the specific accessibility features, if any, that web developers had included on the pages. We found that accessibility violations were prevalent across states but to varying degrees for a specific accessibility criterion. Although violations were detected in all 4 POUR accessibility principles, the most number of known violations occurred in meeting the perceivability and operability principles. Most violations in 508c guidelines occurred in not providing functional text in scripting languages and in not providing text equivalents for non-text. The degree of effort and conformance significantly varied between states; a majority of states exhibited a lower degree of effort, while a few attempted innovative ways to enhance accessibility on their websites. The efforts seemed to focus on meeting the minimum threshold– it is not clear if websites were designed proactively for accessibility.

## 1. Introduction and Background

Since the beginning of the COVID-19 pandemic, every state public health agency in the United States has created and maintained a dedicated website related to the pandemic. These websites have included information on COVID symptoms, guidelines for preventing its spread, dashboards on case counts and trends, information on testing, recommendations for safe reopening of businesses and educational institutions, and more recently, vaccine information. Large swaths of the public, healthcare professionals, including front-line workers and contact tracers, and more recently decision makers and workers in industries, businesses, and educational institutions have come to rely on these websites as a primary daily source of up-to-date information for making lifesaving decisions and as a portal to access and schedule vaccinations for themselves and their families. Given the need to be useful and serve as many people as possible during a pandemic, and to widely communicate and distribute accurate, credible, and life-saving information on COVID-19 and related services such as vaccines, to *all people* in the general population, it is important that these websites be highly *accessible.*

In investigating whether and to what extent COVID-dedicated public health websites have been accessible during this pandemic, a few recently published studies (Alajarmeh, 2021; Alismail & Chipidza, 2021; Dror et al., 2021; Howe et al., 2021; Jo et al., 2021) have examined these websites in the United States and other countries. Four of these five studies have evaluated the extent to which websites violated published and well-accepted web accessibility guidelines, such as the World Wide Web Consortium (W3C) Web Content Accessibility Guidelines (WCAG) (*WCAG Guidelines*, 2022) using automated website accessibility checkers such as AChecker (*AChecker*, 2022) and WAVE checker (*WAVE Accessibility Checker*, 2022). The fifth study by Howe et al. (2021) manually analyzed the viewability of vaccine registration websites, the availability of non-English language options on the pages, and the reading level of information on these pages. The general picture that emerges from these studies is that public health websites, especially those created to disseminate information on COVID-19, violate many accessibility guidelines and principles in general, be it the WCAG guidelines or other specific principles such as the reading levels of information on a page.

 Although these studies provide important preliminary insights into accessibility errors and problems, they are limited in one important respect – they report only aggregated counts of the number and type of known problems, errors, and violations of the broadest WCAG accessibility principles and guidelines (*WCAG Guidelines*, 2022) and decide if a website fails to conform; they do not delve deeper into an analysis of the graduated scales in the conformance levels and success criteria defined for each WCAG guideline. The mere reporting of the number of known problems, errors, and violations, or judging the overall conformity of a website as a "fail" if even one known problem is present at any level of conformance, as the studies using automated checkers cited above have reported, does not reflect the nuance involved in interpreting accessibility violations using graduated scales like the WCAG levels of conformance scale. WCAG organizes its framework in a hierarchy. The highest level in the hierarchy consists of the foundational accessibility principles that websites must be perceivable, operable, understandable, and robust (commonly referred to as the POUR framework). At the next level of the hierarchy, for each principle, WCAG specifies 12 guidelines – these guidelines set goals that web authors should work toward to make content accessible. Finally, at the lowest level of the hierarchy, WCAG provides testable success criteria with three increasing levels of conformance, A (lowest level), AA (medium) and AAA (highest level).  The critical point to note here is that many studies fail to consider when reporting the results that these levels of conformance form an ordinal scale and have a natural order and gradation to them; A is the lowest level of conformance (and hence an easier threshold to meet in practice) than an AAA level of conformance.  Furthermore, not all guidelines specify criteria that can be met at all three levels of conformance: 3 guidelines specify only an A level; 3 others specify A and AA; the remaining 6 guidelines of the 12 specify all three levels of conformance, A, AA, and AAA. Meeting higher levels of conformance would require that the webpage also meet the lower levels of conformance – for example, for a webpage to be classified as conforming to the highest AAA level, it must conform to the AAA, AA, and A criteria specified for that guideline (assuming the guideline has criteria at all three levels).

The main implication of the gradation in the conformance scale is that web designers may choose to aim for a particular level of conformance when they develop a website, particularly given that building accessibility into state department health websites is voluntary, and designers do not have to conform to Federal accessibility standards such as those specified in Section 508c (Section 508 of the Rehabilitation Act of 1973, 1973). If an accessibility guideline requires the designer to meet only the lowest level of conformance of A, it may be easy for the designer to achieve success in conformance. If, on the other hand, a guideline includes all three levels of conformance, a designer may choose to still meet only the lowest level of conformance and not aim for higher levels of conformance. Or they may choose to aim for higher levels of conformance. In judging the overall level of conformance of a website, one must consider violations of accessibility principles *jointly* at all 3 levels (A, AA and AAA) of conformance. For example, for a particular guideline, if a website has no known problems at the highest AAA level, has known problems at the AA level, but has no known problems at the A level, the overall level of conformance of that website would be at the A level. Not having known problems at a certain level for a guideline does not mean that the website conforms at that level. By the same token, having known problems at a certain level does not mean that the website fails to conform; it may just have conformed at a lower level.

Furthermore, given that accessibility guidelines such as the WCAG 2.0 are goals web designers can voluntarily strive to achieve in their development effort, the levels of conformance scale can be reflective of the level of effort web designers may have chosen to apply to successfully conform to accessibility guidelines, which the published studies thus far do not address.

Extending and integrating the ideas that conformance can occur at progressively increasingly difficult levels, and that conformance requires and reflects design and development effort, and further that web developers for state public health agencies are not obligated to adopt WCAG or other guidelines such as Section 508c and do so voluntarily, our main study goal was to evaluate and understand the level of effort in designing for accessibility on health websites dedicated to COVID-19 information in US states. To this end, in addition to using the WCAG 2.0 A, AA and AAA levels of conformance, we extended our overall evaluation of efforts to design for accessibility to include two additional aspects – we included conformance with the Section 508c standard, and we also manually identified, evaluated, and rated the specific accessibility features if any that web developers had included on their website (more than just the website code passing automated checkers). Our rationale was that states did not have to meet Section 508c standards; website developers did not have to provide any additional features for accessibility or inclusion. Therefore, in addition to the WCAG levels of conformance, if a website complied with 508c guidelines as well and if it contained specific visible features that improved accessibility, it would reflect additional efforts to make the website accessible. Furthermore, we sought to address the limitations outlined in previous paragraphs in recently published studies on how overall conformance is interpreted when analyzing reports of accessibility violations with automated checkers. Finally, our objective was to add to existing data on accessibility violations by evaluating the prevalence and pervasiveness of accessibility problems on state health department websites so that we could determine whether a handful of states contributed to the majority of accessibility violations. In particular, we evaluated the home / landing pages and the vaccine information pages in our study, given the importance of these pages during a COVID pandemic.

## 2. Methods
### 2.1. Accessibility evaluations

We used the publicly available and highly validated Web Access Checker tool AChecker (*AChecker*, 2022; Alismail & Chipidza, 2021) to assess accessibility of the 51 public health agency websites in 50 states and Washington, DC. We evaluated violations of both WCAG 2.0 and the Section 508c guidelines using this tool. We selected two different URLs for each public health agency website, their COVID-19 home page, and their vaccine page. We created a spreadsheet with these URLs for public health agency websites in each state to ensure consistency in the URLs we were analyzing. We randomly assigned each state to one of the four researchers (the authors) in the team.

Each researcher used AChecker to identify accessibility violations: three researchers assessed 13 public health agency websites each, while one researcher assessed 12 websites to complete the evaluation of 51 websites. During the evaluation, each researcher entered the homepage URL of the state public health agency website in AChecker and opted for the AAA level WCAG 2.0 as the guideline (Figure 1a) to check against (given this level also reports violations at levels AA and A levels).

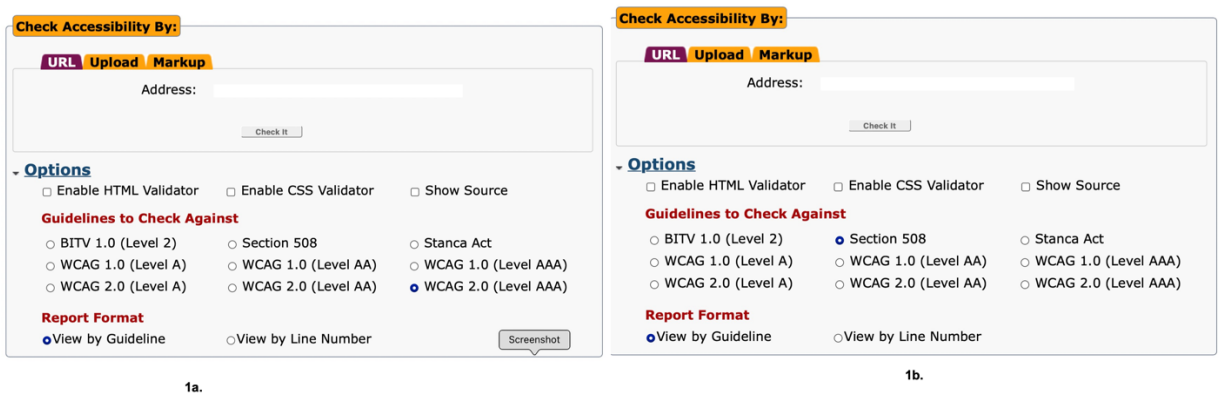

**Figure 1a and b.** AChecker interfaces for checking accessibility of a website using WCAG 2.0 guidelines and Section 508c guidelines. The URL is entered in the Address box in the AChecker.

The A-Checker evaluated these pages and provided a description of the known problems, which we used in our analysis. Once we documented the known problems for the WCAG 2.0 guidelines for the home page, we performed an accessibility check for the 508c guidelines (Figure 1b). After documenting these violations, we repeated the accessibility check for WCAG 2.0 and 508c guidelines for vaccine pages.

In the spreadsheet template we used for the evaluation (Figure 2), for the WCAG 2.0 guidelines, for each state's public health agency website, we documented the link to the website, the corresponding type of page (home page, vaccine page), the number of known problems at each level of the guideline (A, AA, and AAA), the specific guideline item and the criteria that were violated, and any comments from the evaluator (see Figure 2). If more than one type of guideline was violated on a state website, we documented these in separate rows and specifically identified which guideline and criteria were violated.

In the same spreadsheet template, we also documented violations of the 508c guidelines (Figure 2). For the 508c guidelines, we documented the number of known problems under the A, E, I, L, and N guidelines for each type of page (home page, vaccine page) for each state. We also recorded in comment form any concerns that we may have had with a website during the evaluation process.

Once we documented the type and number of known problems provided by the A-checker evaluation for both the WCAG and 508c guidelines, we performed a manual assessment of the home page and the vaccine pages. In particular, we evaluated what functionality or features each state public health website provided for accessibility. For each state, we first visited the URL of the home page and the URL of the vaccine page. Once on that page, we examined the general accessibility features provided on that page.

When examining the pages for functionality supporting accessibility, we were interested in identifying specific features visible on the page such as availability of text resizing tools, and American Sign Language videos, and use of artificial intelligence (AI) based accessibility settings, to mention a few. We documented this information in the spreadsheet template (Figure 2) with open responses to a question we asked for each home page and vaccine page, namely, what does the website (features) do for accessibility?

Figure 2. Spreadsheet template we used to document violations of WCAG 2.0, Section 508c, and unique accessibility functions and features we found on the websites.

## 2.2. Ratings of accessibility support features

Once we documented the accessibility support features that each website offered, we categorized our open-ended comments into four bins based on the type and degree of accessibility support provided on these pages. We devised ratings ranging from 0 to 3 based on the degree of accessibility support that the website offered. See Table 1 for the ratings, what they indicate about the degree of support, and examples of features showing what qualified for a particular rating. Based on this classification and rating scheme, an evaluator rated the accessibility features of each state public health website, followed by a discussion and agreement on the ratings among the three evaluators. We excluded 2 states from this evaluation because the a-checker assessment returned a page-load error. We assessed accessibility features on pages only for 49 states.

| What kinds of features does the website provide for accessibility? | Rating and Bin | Example Features |
|---|---|---|
| None | 0 | - No accessibility links. |
| General information about accessibility policy and "contact us" or the option to report accessibility problems | 1 | - There is a site policy link leading to an accessibility policy page; the policy page provides common shortcuts for browsers; and provides info on related resources; Google language options<br>- Leads to the ADA page https://www.michigan.gov/disabilityresources/0,4563,7-223-74971-372167--,00.html<br>- Provides a page for web accessibility certification, but asks people to contact them to request any alternate formats; provides a text resize tool at the top right. |
| Provides some features such as language selection, contrast control, font size adjustments, American Sign Language videos | 2 | - They had a contrast setting with two options - default and dark - dark was just dark mode. There was also a normal and large text setting button. Underneath, they had a link that said Report Accessibility Issues.<br>- Sign language videos.<br>- The link provides some other resources like the CDC, Disability Hub, etc., which take users to other pages in case of disability needs. Not very organized and usable, but may help.<br>- Deaf Relay: (Hearing or Speech Impaired) 711 or 1-800-735-2942 in the footer<br>- Features language selection, color scheme, font size, and word spacing. |
| Advanced features such as AI support to customize the website according to user needs | 3 | - A sidebar link opens a custom profile setting for adjustments; adjusts the page based on the settings you input; there is also a statement about accessibility; really great implementation; includes all sorts of features - health profiles, content adjustments, color adjustments, and orientation adjustments.<br>- There is a setting button in the top right where the user can turn on accessibility view, as well as enlarge and shrink the font and increase font weight. The site also has an accessibility policy, where they list everything they have done to make the site accessible, even mentioning using the w3.org checker. Additionally, the user can report accessibility issues with the ability to be contacted. This site was designed with accessibility in mind. Best one I have seen so far. |

## 2.3. Data Analysis
## 2.3.1. Analysis of violations of WCAG 2.0 and 508c guidelines

We excluded 2 states from this analysis because the a-checker assessment returned an error. We performed the following analysis only for 49 states.

We collated all known WCAG 2.0 problems for each criteria at A, AA, and AAA levels for each state, for the home pages and vaccine pages separately, providing the finest level of granularity for data analysis. For each criterion at each level (A, AA, or AAA), we calculated the total number of states that had known problems. For example, the number of states that had known problems was computed in criteria 1.4.3 AA. We represented this information in the Pareto charts to provide us with an indication of the prevalence, distribution, and variation of known problems for each criterion, at different levels across states. We also calculated the number of known problems at the guideline level in A, AA, or AA in all states. For example, Guideline 1.2 has three criteria at level A, so the sum of known problems across all three criteria would represent the number of known problems for Guideline 1.2 at level A for that state.

We computed the aggregated conformance counts at the level of the guideline. For example, Guideline 1.2 has 3 criteria at the A level, 2 criteria at the AA level, and 4 criteria at the AAA level. So, a state would conform to level A only if the number of known problems is zero for all the criteria at that level; a state would conform to level AA only if the number of known problems is zero for all criteria at the A level and AA level. Similarly, a state would conform to AAA only if they have zero known problems in each criteria under each level (A, AA, AAA). Note that not all guidelines were represented at all levels. Three guidelines were represented only at the A level, three at the A and AAA levels, and six guidelines were represented at the A, AA and AAA levels. We analyze and represent these conformance and state counts by each guideline at the levels specified in the WCAG to highlight that some of the guidelines do not have all levels represented. This is important in the analysis, given that conformance to a guideline is achieved only if all the criteria for a guideline also conform. For the 508c guidelines, we computed the total number of violations for the criteria A, E, I, L, and N across all states.

### 2.3.2. Conformance analysis relating ratings of accessibility support features and WCAG 2.0 scores

To understand the degree of effort a state public health agency put into its website for the home and vaccine pages, we evaluated and compared the degree to which a state conformed to the WCAG guidelines and the degree to which it provided features to support accessibility. First, we computed conformance rates for each state across A, AA, and AAA levels. We identified the number of possible opportunities that each state had to conform to levels of A, AA, and AAA in all POUR guidelines. At the A level, there were 12 possible opportunities, at the AA level 6 opportunities, and at the AAA level 9 opportunities. Then we computed the number of times a state actually conformed to that level (A, AA, or AAA) across all criteria and guidelines. This represents the raw conformance rates. Given that A is the minimum threshold and AAA was the highest threshold for conformance, we weighted A, AA, and AAA at 1,2 and 3, respectively, and used *raw conformance rate at a level * weight at that level* as the weighted conformance rate. We weighted the raw conformance rates to ensure that the higher degree of effort needed to conform to higher levels of threshold was reflected in the analysis. These weighted conformance percentage rates at each level (A, AA, AAA) represented the level of conformance for the website.

To assess the degree of effort that each state took to make their websites accessible, we represented our rating of accessibility features and degree of conformance at each level in four quadrants representing effort from the least (lower conformance, no, or few features) to the most (higher conformance, better features).

### 3. Results
### 3.1. Violations of WCAG 2.0 POUR principles in homepages and vaccine pages

At the lowest level of conformance A, AChecker results indicate that homepages (figure 3, panel A) and vaccine pages (figure 4, panel A) in 20 states violated the perceivability principle by not providing adequate text alternatives to users for all nontext content). 13 homepages (figure 3, panel B) and 16 vaccine pages (figure 4, panel B) violated the principle of perceivability by not creating content that could be presented to the user in different ways without losing information, structure, and relationships in the content.  21 homepages (figure 3, panel C) and 19 vaccine pages (figure 4, panel C) contained known problems in navigability and, therefore, in fulfilling the principle of operability by helping users determine the purpose of each link in context on the page – the success criteria for making pages navigable evaluates whether each link purpose can be determined by the user from the link text alone or from the text with the link context. Homepages also violated the understandability principle. The web content on four homepages did not identify the language of the page so that user agents (both assistive technologies and conventional user agents) could accurately render and present text and other linguistic content (figure 3, panel D). 19 homepages (figure 3, panel E) and 18 vaccine pages (figure 4, panel D) did not provide labels or instructions to users when the content required user input, thus violating the guideline to provide input assistance to users so that they could avoid and correct mistakes. The homepages of 11 states (figure 3, panel F), and the vaccine pages in 10 states (figure 4, panel E) violated the robustness principle – their content was not robust enough to be parsed and interpreted reliably by a variety of user agents, including assistive technologies.

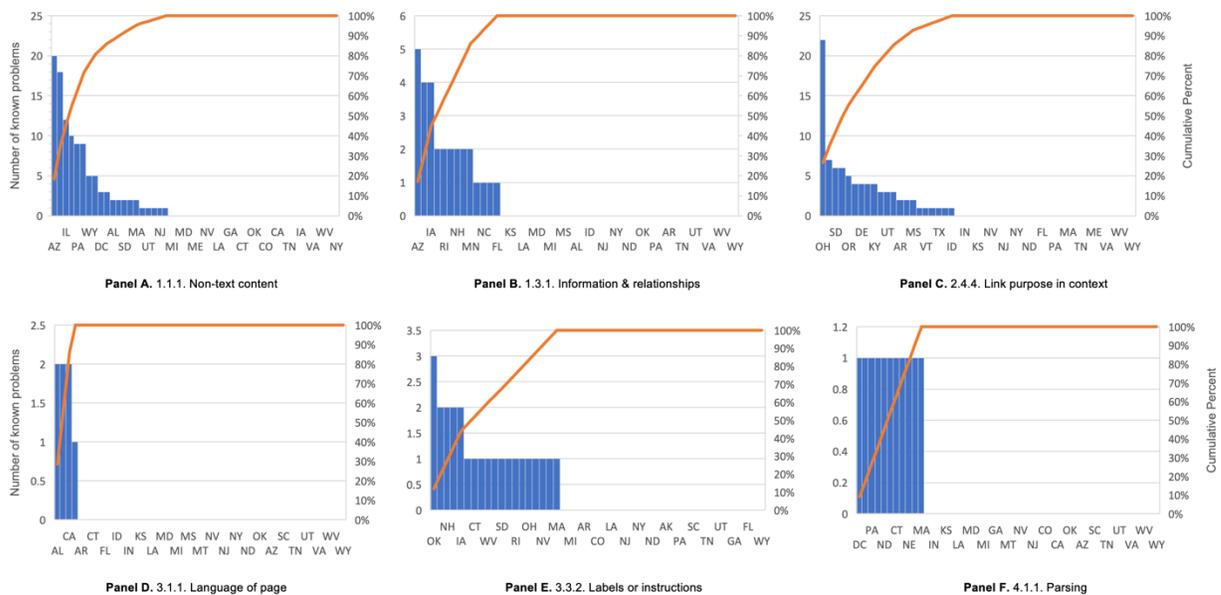

**Figure 3.** The magnitude of known accessibility problems found in homepages of state public health agency websites for minimum A level of conformance. The data are rank ordered by state and separated based on each WCAG 2.0 POUR accessibility principle that was violated and was reported by Achecker to be a known problem at the time of evaluation. Hawaii and Washington are not included as their webpages did not load.

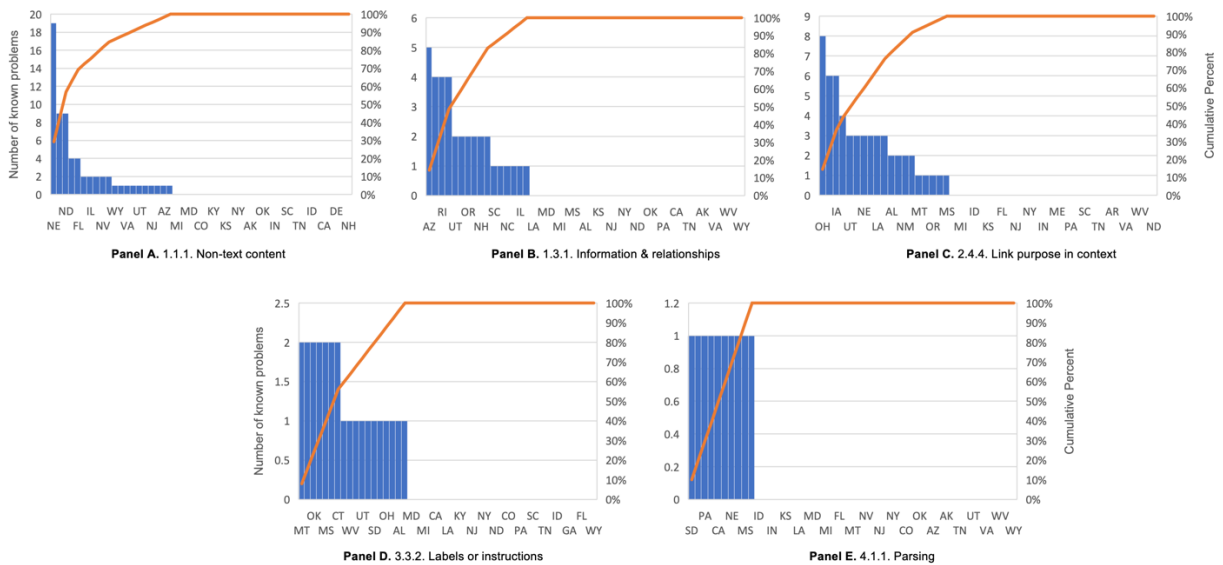

**Figure 4.** The magnitude of known accessibility problems found in vaccine pages of state public health agency websites for minimum A level of conformance. The data are rank ordered by state and separated based on each WCAG 2.0 POUR accessibility principle that was violated and was reported by Achecker to be a known problem at the time of evaluation. Hawaii and Washington are not included as their webpages did not load.

Overall, at the A level of conformance, although all POUR accessibility principles were impacted on both the homepages and the vaccine pages, for each principle, only a few success criteria were violated. Furthermore, for the criteria that were violated, the number of known problems varied from a low of just one problem (figure 3, panel F) to a high of 22 problems (figure 3, panel C) for homepages, and from 1 problem (figure 4, panel E) to a high of 19 problems (figure 4, panel A) for vaccine pages. Most importantly, though, it is clear from the cumulative percentages in panels A through F that only a small number of states contributed to the known problems for these criteria and that the states that contributed to the problems varied for each criteria (the horizontal axis indicates different states in each panel), indicating prevalence of known problems in one criteria or another across states.

Figures 5 and 6 present results of an evaluation of homepages and vaccine pages at the mid-AA level of the WCAG 2.0 conformance criteria. At this level of conformance, 3 home pages (figure 5 panel A) and 1 vaccine page violated the distinguishability guideline by not providing a minimum contrast ratio of at least 4.5: 1 when visually presenting text and images of text. 27 homepages (figure 5 panel B) and vaccine pages (figure 6 panel A) violated the distinguishability guideline by not allowing for up to 20% resizing of text without loss of content or functionality. The violation of this criterion also contributed to the highest number of known problems for the AA conformance level, resulting in 70 problems all in one state for homepages and 62 problems all in one state for the vaccine pages, but also being prevalent in more than half of the 49 states evaluated. Together, poor contrast ratios and inadequate provisions for resizing text, reduced perceivability of the homepages. The homepages in 19 states (figure 5 panel C) and vaccine pages in 16 states (figure 6 panel B) contained known problems in describing the topic or purpose of the pages via the headings and labels they used on the pages. This impacted navigability, which, in turn, resulted in a violation of the operability principle.

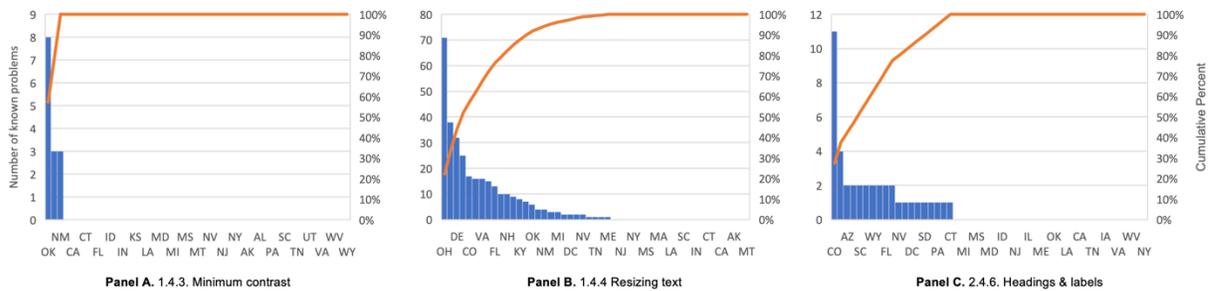

Figure 5. The magnitude of known accessibility problems found in homepages of state public health agency websites for AA level of conformance. The data are rank ordered by state and separated based on each WCAG 2.0 POUR accessibility principle that was violated and was reported by Achecker to be a known problem at the time of evaluation. Hawaii and Washington are not included as their webpages did not load.

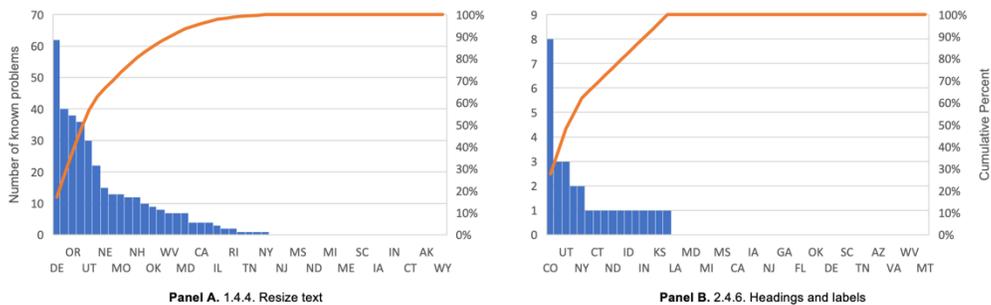

Figure 6. The magnitude of known accessibility problems found in vaccine pages of state public health agency websites for AA level of conformance. The data are rank ordered by state and separated based on each WCAG 2.0 POUR accessibility principle that was violated and was reported by Achecker to be a known problem at the time of evaluation. Hawaii and Washington are not included as their webpages did not load.

Overall, at the AA level of conformance, we saw a similar trend to the A level of conformance in that only a few principles and success criteria were violated. The number of known problems at this level of conformance ranged from a low of 8 (figure 5 panel A) to a high of 71 (figure 5 panel B) for homepages and from a low of 8 (figure 6 panel B) to a high of 62 (figure 6 panel A). Although the number of states that contributed to the known problems varied between a low of just 3 states to a high of 27 states (figure 5 panel B) for homepages and between a low of just one state to a high of 27 states, we observed again that a significant percentage of the known problems came only from a handful of states and that the states contributing to the problems varied for each criteria at the AA level of conformance.

Figures 7 and 8 present the results of an evaluation of homepages at the highest AAA level of the WCAG 2.0 conformance criteria. At this highest level of conformance, 20 homepages and 17 vaccine pages had known problems providing an enhanced contrast of at least 7:1 when visually presenting text and images of text, thus violating the distinguishability guideline and reducing the perceivability of their homepages. This criterion also resulted in the highest number of known problems, 107 in 2 states and over 50 in a few more states for home pages and 128 in 2 states and over 40 in a few more states for vaccine pages. Interestingly, at the AAA level of conformance, even fewer states contributed to a larger cumulative percentage of known problems than with the A or AA levels of conformance. Furthermore, the homepages that contained known problems at the AAA level were different from those at the A or AA levels.

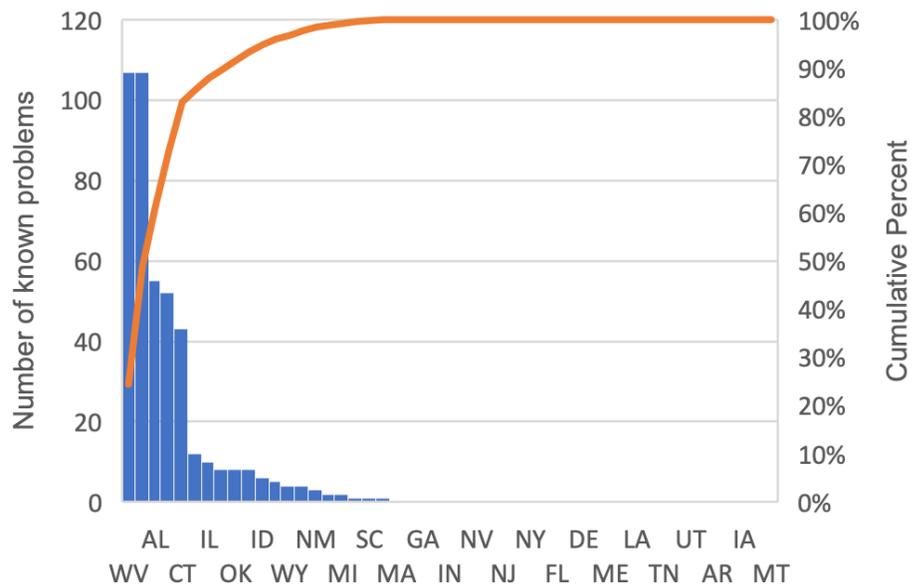

**Panel A.** 1.4.6. Enhanced Contrast

**Figure 7.** The magnitude of known accessibility problems found in homepages of state public health agency websites for the highest AAA level of conformance. The data are rank ordered by state and separated based on each WCAG 2.0 POUR accessibility principle that was violated and was reported by Achecker to be a known problem at the time of evaluation. Hawaii and Washington are not included as their webpages did not load.

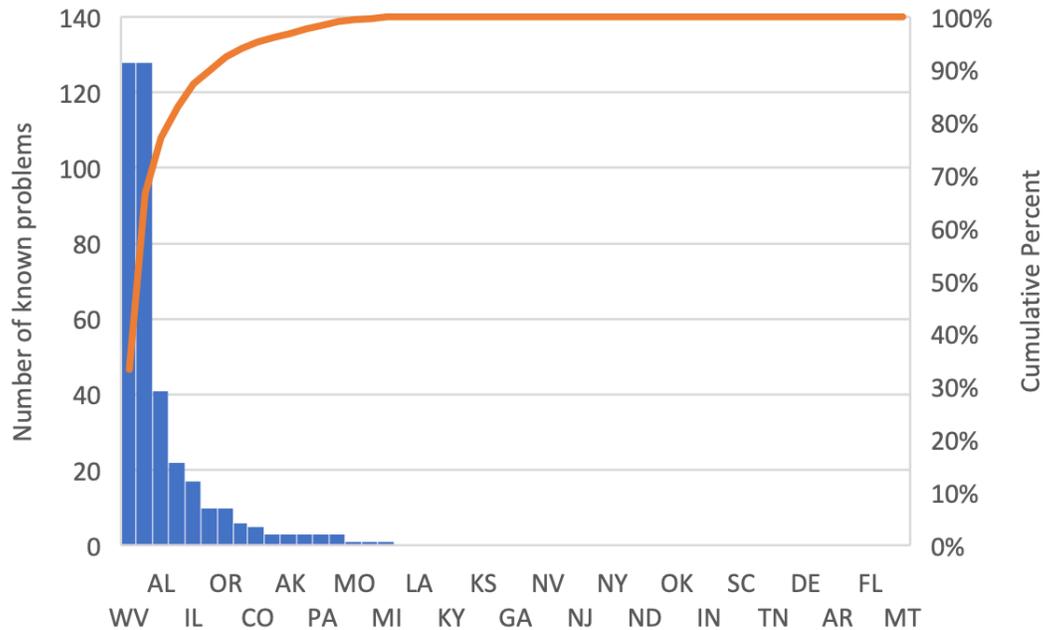

**Panel A.** 1.4.6. Enhanced contrast

**Figure 8.** The magnitude of known accessibility problems found in vaccine pages of state public health agency websites for the highest AAA level of conformance. The data are rank ordered by state and separated based on each WCAG 2.0 POUR accessibility principle that was violated and was reported by Achecker to be a known problem at the time of evaluation. Hawaii and Washington are not included as their webpages did not load.

When the WCAG 2.0 guideline specified that the developer meet only the lowest level of the A threshold of success criteria, the homepages (Table 2) and the vaccine pages (Table 3) in all states had difficulty meeting even the lowest thresholds. Three WCAG 2.0 guidelines specified success criteria at both minimum level A and highest level AAA (Table 4 for homepages and Table 5 for vaccine pages). These guidelines not only had the fewest known problems (only 4 at the A level for Guideline 2.1), but also met the highest levels of conformance.

**Table 2. Known problems in homepages across all states for guidelines that only specify an A level of conformance.**

| Guideline | Number of known problems |
|---|---|
| 1.1. Providing text alternatives | 109 |
| 1.3. Creating adaptable content that can be presented in different without losing information or structure | 29 |
| 4.1. Maximizing compatibility with current and future user agents, including assistive technologies | 11 |

**Table 3. Known problems in vaccine pages across all states for guidelines that only specify an A level of conformance.**

| Guideline | Number of known problems |
|---|---|
| 1.1. Providing text alternatives | 65 |
| 1.3. Creating adaptable content that can be presented in different without losing information or structure | 35 |
| 4.1. Maximizing compatibility with current and future user agents, including assistive technologies | 10 |

**Table 4. Known problems in homepages across all states for guidelines that specify both A and AAA levels of conformance.**

| Guideline | Known problems at the A level | Known problems at the AAA level |
|---|---|---|
| 2.1. Making all functionality keyboard accessible | 4 | 0 |
| 2.2. Providing users enough time to read and use the content | 0 | 0 |
| 2.3. Designing content to avoid seizures in users (such as use of flashes) | 0 | 0 |

**Table 5. Known problems in vaccine pages across all states for guidelines that specify both A and AAA levels of conformance.**

| Guideline | Known problems at the A level | Known problems at the AAA level |
|---|---|---|
| 2.1. Making all functionality keyboard accessible | 4 | 0 |
| 2.2. Providing users enough time to read and use the content | 0 | 0 |

| 2.3. Designing content to avoid seizures in users (such as use of flashes) | 0 | 0 |

Six WCAG guidelines specified success criteria at all three levels of conformance, A, AA, and AAA (Table 6 for home pages, Table 7 for vaccine pages). Interestingly, when web developers could aim for all 3 levels of conformance, while homepages and vaccine pages satisfied guideline 1.4 at the lowest A level, they contained the highest numbers of known problems at the AA and AAA levels – specifically, problems with contrasts in text and images, and inability to resize the test on the homepage contributed to these increased violations of success criteria.

In providing ways to help users navigate the content on the homepages and vaccine pages, the developers satisfied the highest AAA levels of conformance with no known problems but violated the A and AA levels of conformance with 84 and 40 known problems on the homepages and 58 and 29 known problems on the vaccine pages. We also observed this pattern of violation in the guidelines meant to make text content readable and understandable and, more importantly, to help users avoid and correct errors.

Table 6. Known problems in homepages across all states for guidelines that specify A, AA and AAA levels of conformance.

| Guideline | Number of known problems at the A level | Number of known problems at the AA level | Number of known problems at the AAA level |
|---|---|---|---|
| 1.2. Providing alternatives for time-based media | 0 | 0 | 0 |
| 1.4. Making content distinguishable so users can see and hear content including by separating foreground from background | 0 | 333 | 439 |
| 2.4. Providing ways to help users navigate, find content and determine where they are | 84 | 40 | 0 |
| 3.1. Making text content readable and understandable | 7 | 0 | 0 |
| 3.2. Making web pages appear and operate in predictable ways | 0 | 0 | 0 |
| 3.3. Helping users avoid and correct mistakes | 25 | 0 | 0 |

**Table 7. Known problems in vaccine pages across all states for guidelines that specify A, AA and AAA levels of conformance.**

| Guideline | Number of known problems at the A level | Number of known problems at the AA level | Number of known problems at the AAA level |
|---|---|---|---|
| 1.2. Providing alternatives for time-based media | 0 | 0 | 0 |
| 1.4. Making content distinguishable so users can see and hear content including by separating foreground from background | 0 | 365 | 385 |
| 2.4. Providing ways to help users navigate, find content and determine where they are | 58 | 29 | 0 |
| 3.1. Making text content readable and understandable | 5 | 0 | 0 |
| 3.2. Making web pages appear and operate in predictable ways | 0 | 0 | 0 |
| 3.3. Helping users avoid and correct mistakes | 25 | 0 | 0 |

For guidelines that specified only the lowest conformance threshold (only A level), many state homepages (Table 8) and 10 to 20 vaccine pages (Table 9) had difficulty meeting the success criteria, particularly providing text alternatives on their homepages. For guidelines that specified the lowest levels of A and the highest levels of AAA conformance, nearly all states met the AAA criteria (and therefore also met the A criteria) for both homepages (Table 10) and vaccine pages (Table 11). For guidelines that included all three levels of conformance criteria (A, AA and AAA), homepages did progressively worse at the AA and AAA levels of conformance in making the homepage content distinguishable – only home pages of 16 states (Table 12), and vaccine pages of 17 states (Table 13) conformed at the highest AAA level (and thus levels AA and A as well for this guideline. Almost half of the homepages we evaluated had difficulty providing ways for users to navigate their pages – this non-conformance was particularly pronounced at the AA and AAA levels. The only other guideline to which 19 state homepages and 18 state vaccine pages had trouble complying at all three levels of conformance was to help users avoid and correct mistakes.

**Table 8. Number of state public health homepage conformances when the guideline specified only an A level of conformance. Conformance at the A level only requires success in fulfilling A level criteria.**

| Guideline | Number of state homepages with A level conformance (passing only A criteria) |
|---|---|
| 1.1. Providing text alternatives | 29 |
| 1.3. Creating adaptable content that can be presented in different ways without losing information or structure | 36 |
| 4.1. Maximizing compatibility with current and future user agents, including assistive technologies | 38 |

**Table 9. Number of state public health vaccine page conformances when the guideline specified only an A level of conformance. Conformance at the A level only requires success in fulfilling A level criteria.**

| Guideline | Number of state vaccine pages with A level conformance (passing only A criteria) |
|---|---|
| 1.1. Providing text alternatives | 29 |
| 1.3. Creating adaptable content that can be presented in different ways without losing information or structure | 33 |
| 4.1. Maximizing compatibility with current and future user agents, including assistive technologies | 39 |

**Table 10. Number of state public health homepage conformances when the guideline specified A and AAA levels of conformance. Conformance at the A level only requires success in fulfilling A level criteria; Conformance at the AAA level requires success in fulfilling the AAA and the A level criteria.**

| Guideline | Number of state homepages with A level conformance (passing A criteria) | Number of state homepages with AAA level conformance (passing A and AAA criteria) |
|---|---|---|
| 2.1. Making all functionality keyboard accessible | 47 | 47 |
| 2.2. Providing users enough time to read and use the content | 49 | 49 |
| 2.3. Designing content to avoid seizures in users (such as use of flashes) | 49 | 49 |

Table 11. Number of state public health vaccine page conformances when the guideline specified A and AAA levels of conformance. Conformance at the A level only requires success in fulfilling A level criteria; Conformance at the AAA level requires success in fulfilling the AAA and the A level criteria.

| Guideline | Number of state vaccine pages with A level conformance (passing A criteria) | Number of state vaccine pages with AAA level conformance (passing A and AAA criteria) |
|---|---|---|
| 2.1. Make all functionality keyboard accessible | 47 | 47 |
| 2.2. Providing users enough time to read and use the content | 49 | 49 |
| 2.3. Designing content to avoid seizures in users (such as use of flashes) | 49 | 49 |

Table 12. Number of state public health homepage conformances when the guideline specified an A, AA and AAA levels of conformances. Conformance at the A level only requires success in fulfilling A level criteria; conformance at the AA levels requires success in fulfilling both A and AA level criteria for the guideline; conformance at the AAA level requires success in fulfilling A, AA and AAA criteria.

| Guideline | Number of state homepages with A level of conformance (passing A criteria) | Number of state homepages with AA level of conformance (passing A and AA criteria) | Number of state homepages with AAA level conformance (passing A, AA, AAA criteria) |
|---|---|---|---|
| 1.2. Providing alternatives for time-based media | 49 | 49 | 49 |
| 1.4. Making content distinguishable so users can see and hear content including by separating foreground from background | 49 | 22 | 16 |

| Guideline | | | |
|---|---|---|---|
| 2.4. Providing ways to help users navigate, find content and determine where they are | 27 | 18 | 18 |
| 3.1. Making text content readable and understandable | 45 | 45 | 45 |
| 3.2. Making web pages appear and operate in predictable ways | 49 | 49 | 49 |
| 3.3. Helping users avoid and correct mistakes | 30 | 30 | 30 |

Table 13. Number of state public health vaccine page conformances when the guideline specified an A, AA and AAA levels of conformances. Conformance at the A level only requires success in fulfilling A level criteria; conformance at the AA levels requires success in fulfilling both A and AA level criteria for the guideline; conformance at the AAA level requires success in fulfilling A, AA and AAA criteria.

| Guideline | Number of state vaccine pages with A level of conformance (passing A criteria) | Number of state vaccine pages with AA level of conformance (passing A and AA criteria) | Number of state vaccine pages with AAA level of conformance (passing A, AA, AAA criteria) |
|---|---|---|---|
| 1.2. Providing alternatives for time-based media | 49 | 49 | 49 |
| 1.4. Making content distinguishable so users can see and hear content including by separating foreground from background | 49 | 22 | 17 |
| 2.4. Providing ways to help users navigate, find content and determine where they are | 28 | 17 | 17 |
| 3.1. Making text content readable and understandable | 46 | 46 | 46 |
| 3.2. Making web pages appear and operate in predictable ways | 49 | 49 | 49 |
| 3.3. Helping users avoid and correct mistakes | 31 | 31 | 31 |

## 3.2. Violation of Section 508c guidelines

Figures 9 and Figure 10 present results that evaluate the homepages for compliance with the 508c accessibility standards. The results indicate that 21 out of the 49 state homepages (figure 9 panel A) and 17 out of the 49 (figure 10 panel A). state vaccine pages evaluated did not comply with providing text equivalents for nontext content for a total of 82 noncompliances in homepages and 25 noncompliances for vaccine pages.

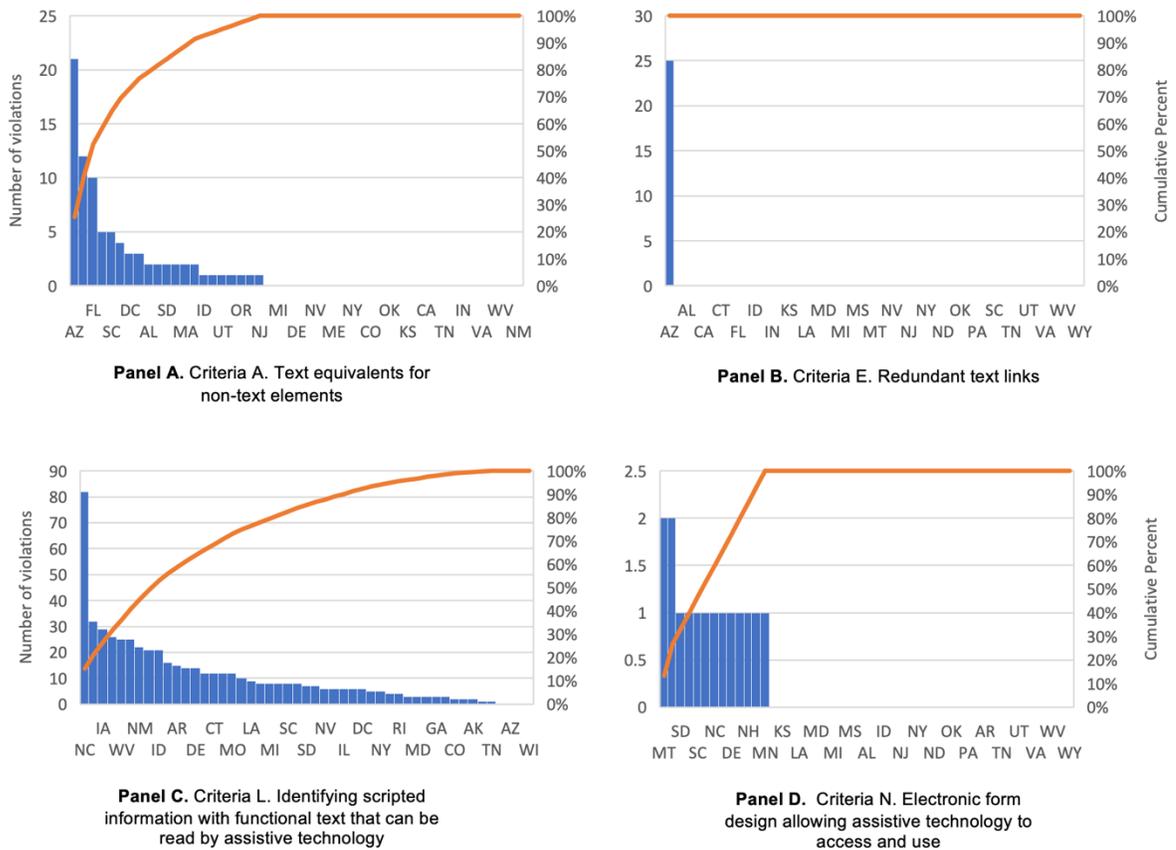

**Figure 9.** The magnitude of accessibility violations of the 508c standards found in homepages of state public health agency websites. Hawaii and Washington are not included as their webpages did not load.

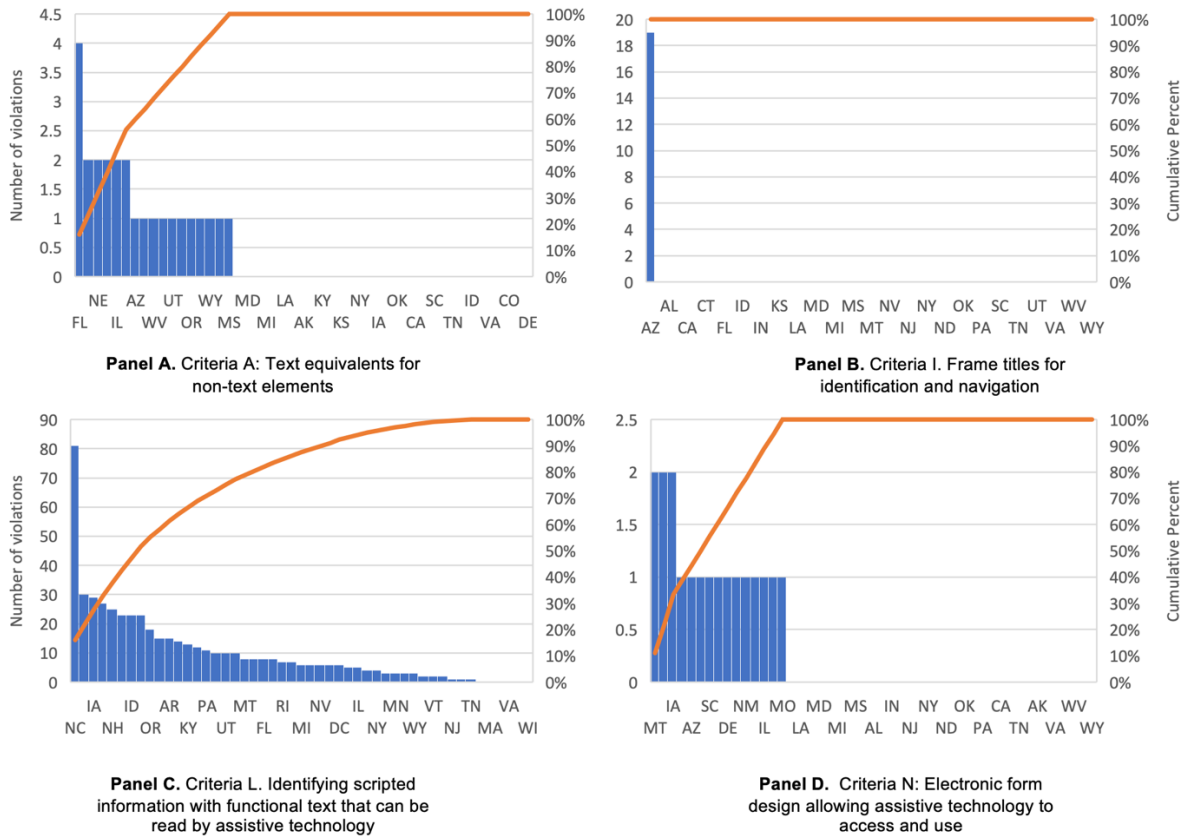

**Figure 10.** The magnitude of accessibility violations of the 508c standards found in vaccine pages of state public health agency websites Hawaii and Washington are not included as their webpages did not load.

One state accounted for all 25 violations on home pages (figure 9 panel B) in not providing redundant text links for active regions of server-side image maps, and 19 violations on vaccine pages (figure 10 panel B) in not providing frame titles on the pages so identification and navigation of the frames on the vaccine pages could be easy for the user.

The highest number of violations and the most prevalent was with the lack of functional text in the scripting language that assistive technology could read – the homepages of 45 states (figure 9 panel C) and vaccine pages of 43 states (figure 10 panel C) contained 534 violations and 504 violations of this criterion respectively. Only a handful of state home pages (13) (figure 9 panel D) and state vaccine pages (15) (figure 10 panel D) did not meet the criterion of making forms readable by assistive technology.

### 3.3. Level of WCAG 2.0 conformance and accessibility feature rating

When we rated the degree of accessibility features and functions contained in the homepages based on their level of conformance to the WCAG 2.0 criteria, across all levels of conformance (A, AA and AAA), only one homepage (Figure 11) and none of the vaccine pages (Figure 12) met the highest levels of conformance (100%) and received our highest rating (a 3 rating) for the best accessibility features and functions on the page. Three homepages and one vaccine page achieved the highest degree of conformance (100% at the AAA level) and also received our rating for enhanced features. Five homepages and vaccine pages had 100% conformance at the AAA level, but our rating indicated lack of or very few features to improve accessibility or inclusion. 25 homepages and 26 vaccine pages had AAA conformance levels between 89% and 44%, and between 89% and 56%, respectively, and our rating indicated availability of very few features. 16 homepages and 17 vaccine pages had conformance levels between 89% and 67% and our rating indicated availability of features to enhance accessibility. At the AAA level, only one homepage and none of the vaccine pages had a conformance below 50% that our rating also indicated lack of accessibility features.

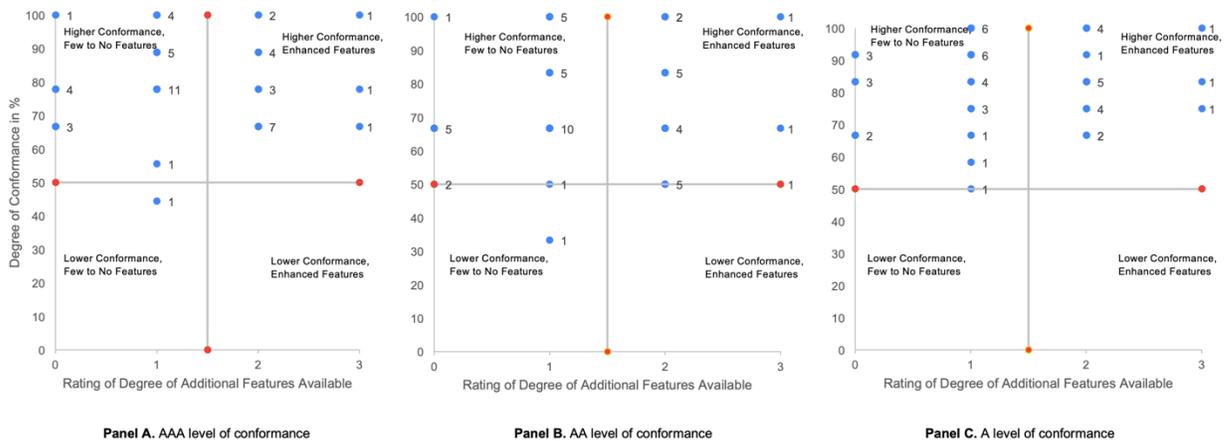

**Figure 11.** Degree of conformance to accessibility principles versus analyst rating of the degree of accessibility and inclusivity features provided on homepages.

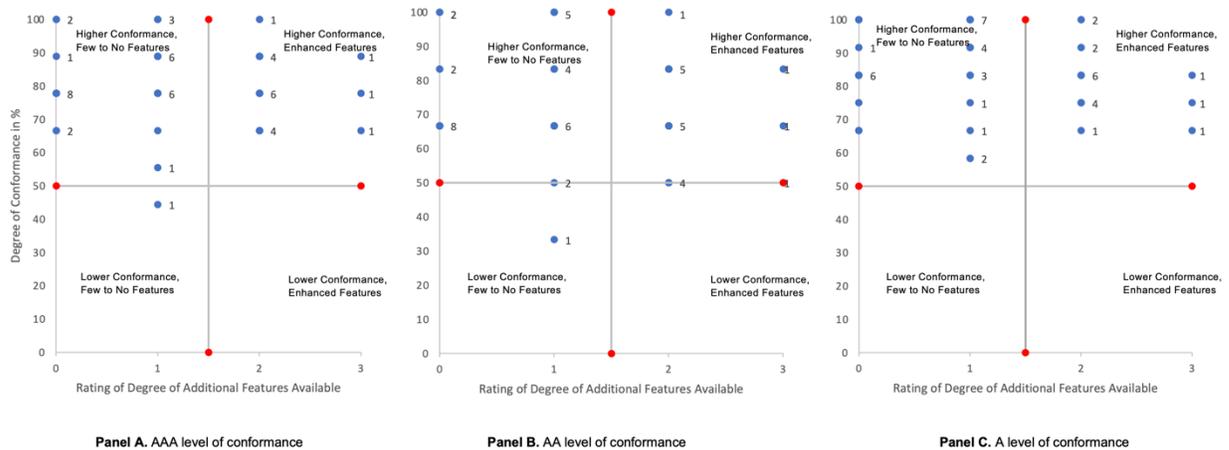

**Panel A.** AAA level of conformance  **Panel B.** AA level of conformance  **Panel C.** A level of conformance

**Figure 12.** Degree of conformance to accessibility principles versus analyst rating of the degree of accessibility and inclusivity features provided on vaccine pages.

Similar patterns were seen in the conformance in levels AA and A and features provided, with the majority of the states providing little to no visible accessibility features when they had higher conformance rates, and some having higher conformance rates and accessibility and inclusivity features. Three homepages and one vaccine page were 100% conformant to the AA level and our rating indicated availability of enhanced features. 6 homepages and 7 vaccine pages had 100% conformance at the AA level but our rating indicated lack of any accessibility features. 24 homepages and vaccine pages had conformance levels between 83% and 33% and our rating indicated lack of features. Of those 24 homepages and vaccine pages, 4 pages had conformance levels of 50% or less and our rating indicated lack of accessibility features. 16 homepages and 17 vaccine pages had conformance levels ranging from 83% to 50% but our rating indicated availability of accessibility features.

At the A level of conformance, 5 homepages and 2 vaccine pages had 100% conformance levels and our rating indicated availability of accessibility features, while 6 homepages and 10 vaccine pages had 100% conformance but our rating indicated lack of features. 24 homepages and 21 vaccine pages had conformance levels ranging from 92% to 50% and from 92% to 58%, respectively, and our rating indicated lack of any features. Of these 24 homepages, only one page had a conformance level of 50%.

14 homepages and vaccine pages had conformance levels between 92% and 67% but our rating indicated availability of accessibility features.

## 4. Discussion

The main objective of our study was to evaluate, on a spectrum of increasing levels of accessibility compliance requirements, the results of the efforts made by website developers to incorporate and meet accessibility compliance criteria on websites designed for state public health departments during the pandemic.

The results of our 3 sets of evaluations point to a few major themes about the accessibility of landing pages and vaccine pages of public health department websites in US states: (1) although violations were detected in all 4 POUR accessibility principles to varying degrees across all states, overall, in the homepages and in the vaccine pages, the most number of known violations occurred in meeting the perceivability and operability principles; (2) violations were prevalent across all states with known problems in conforming with one criteria or another, homepages only in 5 states conformed 100% to all principles and criteria without any known problems according to the WCAG 2.0 principles; vaccine pages only in 3 states conformed 100% (3) not all states violated the same criteria; for specific principles, only a handful of states contributed to the violations; (4) most violations in 508c guidelines occurred in not providing functional text in scripting languages to enable assistive technologies to read the text and in not providing text equivalents for non-text closely matching the violations occurring in the WCAG 2.0 criteria (5) we can conclude that the effort in designing for accessibility seemed to focus on meeting the WCAG guidelines and in being compliant without errors – it is not clear though, if they proactively designed for accessibility, or if in their design process, they iteratively checked for errors during the website build process.

Our overall evaluation and finding that, in general, the accessibility of public health websites leaves something to be desired and that they require more work and effort to bring them to conform with accessibility standards are consistent with the general sentiment expressed in other recent studies that have investigated accessibility, particularly accessibility of health-related websites (Acosta-Vargas et al., 2020; Alismail & Chipidza, 2021; Dror et al., 2021; Howe et al., 2021; Jo et al., 2021).

Our specific evaluations using the automated AChecker indicating that state public health websites violated the WCAG 2.0 criteria also echo findings from other similar studies (Alismail & Chipidza, 2021; Jo et al., 2021). Our findings closely echo Jo et al. (2021) in websites having most known problems from poor contrast and poor alternative text options.

Two larger studies, one by Dror et al. (2021) and the other by Alajarmeh (2021), examined healthcare websites around the world and evaluated their conformity to the WAI and WCAG guidelines. Dror et al. (2021) concluded that robustness (39%) and perceptibility (32%) were impacted more than operability (19%) and understandability (10%). Although their findings are not directly comparable to ours because they evaluate a different set of websites at different points in time, our findings that perceptibility and operability were violated to a greater degree than robustness are in contrast to theirs. This may point to typical accessibility principles that web developers generally have trouble with.

Our findings, however, closely match Alajarmeh (2021) in that they also found that although all WCAG POUR principles were violated in most of the websites, the problems detected mostly related to the perception of information and the operability of the interface items. In particular, the top recurring errors related to the lack of alternative descriptive text for interface items such as links and images, and potential incompatibilities with assistive technologies, findings that were also echoed by our 508c evaluation. Both the WCAG and 508c violations together point to a significant concern in enabling accessibility of nontext content and providing better support for assistive technologies.

We found the discrepancies between conformance rates and our ratings of accessibility features interesting. While some states conformed 100%, only a few states also received our highest rating indicating what was visible and perceptible on a page for a user, and what was coded into the page for compliance, may not necessarily always match. This also raises questions of whether websites that conformed 100% enhanced experiences of users with disabilities compared to sites that had 100% conformance and included features that were visible on the page. Furthermore, some pages received our highest rating for features, but did not conform 100%. Few of these pages had AI enabled accessibility modules – these pages, surprisingly, did not conform to the WCAG guidelines, because we think the AChecker may not have captured the AI module as an integrated part of the webpage.

One major reason websites continue to be designed in an inaccessible way may be related to the knowledge of website programmers and developers about accessibility design guidelines. Lazar et al. (2004) in their seminal study of webmasters found that webmasters were knowledgeable about WCAG guidelines. But recent research literature both among web developers and developers-in-training is replete with studies that conclude that they do not keep up with the many nuances involved in interpreting and implementing website accessibility guidelines and criteria such as the WCAG (Antonelli et al., 2018; Ballesteros et al., 2015; Durdu & Yerlikaya, 2020; Farrelly, 2011; Ferati & Vogel, 2020; Freire et al., 2008; Harper & Chen, 2012; Inal et al., 2019, 2020; Inal & Ismailova, 2018; Lazar & Greenidge, 2006; Lopes et al., 2010; Pichiliani & Pizzolato, 2019; Sánchez-Gordón & Moreno, 2014; Sørum et al., 2013; Tigwell et al., 2017). Of particular importance and concern is the recent work by Pichiliani & Pizzolato (2021), who found that an overlooked and critical consideration is a lack of understanding among web developers of cognitive disabilities and the science behind these disabilities, resulting in the absence of specific guidelines and web contexts that may be impacted by these cognitive disabilities.

Furthermore, studies show that there may be poor understanding and use of the authoring and development tools and technologies used for ensuring accessible web designs (Freire et al., 2007, 2008; Harper & Chen, 2012; Lazar et al., 2004; Miñón et al., 2014; Power et al., 2012), an overreliance on automation during coding(Inal et al., 2019; Jaeger, 2008; Law et al., 2005; Spyridonis & Daylamani-Zad, 2021), lack of tools to assess whether their design choices such as color would make it accessible to the visually disabled (Tigwell et al., 2017), and in general, a paucity of studies directly interviewing web developers to understand their processes (Holliday, 2020) making design of accessible websites challenging. The prevalence of accessibility problems across states found in our study is indicative of potential gaps in knowledge and use of accessibility guidelines and authoring tools, but it also raises questions on the effectiveness of guidelines and the availability of easy-to-use authoring tools for rapid website proofing and production.

Part of the reason for a widespread lack of awareness of accessibility guidelines may be an inadequate curriculum for website developers including availability of dedicated courses (Ferati & Vogel, 2020; Harper & Chen, 2012; Katsanos et al., 2012; Youngblood et al., 2018) and lack of appropriate training including in providing clarity in the differences between usability, accessibility, and user experience (Kamoun & Basel Almourad, 2014; Keates & Clarkson, 2003; Powlik & Karshmer, 2002; Sauer et al., 2020; Stephanidis & Savidis, 2001). Being educators and researchers in human factors engineering ourselves, we agree with other researchers that the path from being a student developer to a full-fledged practicing web developer also needs to be revisited.

Many resource constraints when designing a website including budget constraints, upfront costs (Antonelli et al., 2018; Farrelly, 2011; Inal et al., 2019; Lazar et al., 2004; Trewin et al., 2010) and lack of adequate buy-in (Harper & Chen, 2012; Inal et al., 2019, 2020; Lazar et al., 2004; Putnam et al., 2012), seem to impact whether and to what extent website designers consider accessibility during their design. But researchers also point out that fears that designing websites for accessibility may be expensive and will take enormous effort may be unfounded (Brady & Bigham, 2015; Harper & Chen, 2012; Law et al., 2005; Loiacono & Djamasbi, 2013; Pichiliani & Pizzolato, 2021). Given the rapid spread of COVID-19 during the initial phases of the pandemic, we suspect that state public health agencies were under enormous time pressure to get websites up and running quickly, and to disseminate vital health information infection risk and how to minimize this risk. Therefore, conforming to the highest accessibility standards may have taken a bit of a backseat compared to rapidly disseminating web content with lifesaving information about a rapidly spreading disease.

Indeed, there is support in the literature (Lazar et al., 2004) for the idea that some web developers tend to make websites accessible only when this is a legal requirement or only when a customer or a client request that the sites be accessible. However, recent studies suggest that webmaster attitudes have changed tremendously from Lazar's seminal study (2004), so much so that many web developers now consider it their ethical obligation to make websites accessible (Antonelli et al., 2018; Farrelly, 2011; Freire et al., 2008; Inal et al., 2019, 2020; Pichiliani & Pizzolato, 2019; Putnam et al., 2012; Schmutz et al., 2016). In the case of COVID-19 dedicated state public health websites, it is not clear whether state agencies requested the vendors to design the sites to conform to the highest standards of accessibility possible or whether the web development vendors did not do so because of the added costs and time needed. This is especially reflected in our findings on the varying degrees of effort applied to improve accessibility features of the website. While some websites incorporated accessibility features, others only aimed to conform.

Researchers also believe that there is a gap between knowledge of accessibility standards and translating that knowledge into implementation and practice. Many models of technology adoption and implementation have been developed in the literature, such as the diffusion of innovation model (Rogers, 2003), the technology acceptance model (Davis et al., 1989), the theory of reasoned action (Ajzen & Fishbein, 1980) the theory of planned behavior (Ajzen, 1991; Vollenwyder et al., 2019) and the unified theory of acceptance of use of technology (Venkatesh et al., 2003). Specific accessibility models like the causal loop diagram (Abdelgawad et al., 2010), WIAM Web accessibility integration model (Lazar et al., 2004), the process model for continuous testing of web accessibility (Campoverde-Molina et al., 2021), and machine learning and statistical tools or evaluating accessibility (Lundgard & Satyanarayan, 2021; Sutton, 2020; Yu & Parmanto, 2011) also have been developed, but the extent to which these models have changed day-to-day website design and development practices needs a deeper evaluation. It is surprising that, although 31 years have passed since the time the first ever website went live in 1991, and nearly 22 years have passed since the first WCAG guidelines were formalized and published in 1999, and all the development in web tools and technologies that have occurred in these years, websites continue to have problems with accessibility and that widespread awareness and knowledge and awareness of website accessibility guidelines and standards may be lacking, especially among website programmers and developers. Indeed, what may be needed is to view accessibility design not just as satisfying guidelines and complying with criteria (Farrelly, 2011), but viewing accessibility holistically in the broader context of equity and inclusion of broader populations (Holliday, 2020; Kelly et al., 2004; Sloan et al., 2006).

## 5. Study limitations

We conducted automated accessibility checks and manual evaluation of accessibility features, but we did not use assistive technologies or conduct user testing with people with disabilities to understand how they experience these websites. This is an important consideration for future studies. We also did not engage designers to understand their design processes and barriers in their design process for creating accessible websites during a pandemic. Finally, it might be beneficial to understand the cost and resources allocated to design efforts for accessibility to better design future websites.

## 6. Conclusions

The main objectives of our study were to understand the accessibility of state public health websites during COVID-19 given the crucial role that websites play in providing public health information on COVID-19 guidance and prevention, and given the disparities already faced by people with disabilities due to the pandemic. We evaluated the prevalence and variability of accessibility problems and the criteria violated, and the degree of effort that state public health agency websites demonstrated to provide or enhance accessible information on home and vaccine pages for 51 states. We found that accessibility violations were prevalent across states but to varying degrees for a specific accessibility criterion. Our evaluation indicates that state public health websites can improve support for nontext content, use of assistive technologies, make their content distinguishable, and improve navigation to support website users who need accessible content. The degree of effort and conformance significantly varied between states; a majority of states exhibited a lower degree of effort, while a few attempted innovative ways to enhance accessibility on their websites.

We found that a few public health websites used innovative AI-based, floating accessibility modules that were layered on the webpage that users could customize and turn on and off depending upon their specific disabilities. AI technology holds immense potential for not only making websites accessible for users at the click of a button, but it can also reduce web page authoring and coding burdens for designers, particularly when important accessibility considerations such as alternative text for images, or contrast, color and font changes have to be implemented – these do not have to be coded into the webpages.

Identifying and addressing factors that promote accessibility of information during public health crises with a holistic understanding of website developer knowledge, availability, usability, and reliability of authoring support and evaluation tools, awareness and knowledge of disability science, and organizational culture and policies is increasingly needed in the digital age to have a more equitable, diverse, and inclusive public digital space.

**Conflict of Interest:** Priyadarshini R. Pennathur is a scientific editor in this journal. She was not involved in any of the steps in the peer review process for this manuscript.

Sutton, H. (2020). Accessible COVID-19 tracker enables a way for visually impaired to stay up to date. *Disability Compliance for Higher Education*, *25*(11), 9–9. https://doi.org/10.1002/dhe.30856

Tigwell, G. W., Flatla, D. R., & Archibald, N. D. (2017). ACE: a colour palette design tool for balancing aesthetics and accessibility. *ACM Transactions on Accessible Computing (TACCESS)*, *9*(2), 1–32.

Trewin, S., Cragun, B., Swart, C., Brezin, J., & Richards, J. (2010a). Accessibility challenges and tool features: An IBM Web developer perspective. *Proceedings of the 2010 International Cross Disciplinary Conference on Web Accessibility (W4A)*, 1–10.

*Universal Design: What is it?| Section508.gov*. (2021). http://section508.gov/blog/Universal-Design-What-is-it

Section 508 of the Rehabilitation Act of 1973, (1973). https://www.section508.gov/manage/laws-and-policies/

Venkatesh, Morris, Davis, & Davis. (2003). User Acceptance of Information Technology: Toward a Unified View. *MIS Quarterly*, *27*(3), 425. https://doi.org/10.2307/30036540

Vollenwyder, B., Iten, G. H., Brühlmann, F., Opwis, K., & Mekler, E. D. (2019). Salient beliefs influencing the intention to consider Web Accessibility. *Computers in Human Behavior*, *92*, 352–360. https://doi.org/10.1016/j.chb.2018.11.016

*WAVE Accessibility Checker*. (2022). https://wave.webaim.org

*WCAG Guidelines*. (2022). https://www.w3.org/WAI/standards-guidelines/wcag/
43